\documentclass[a4paper, twocolumn, 10pt, superscriptaddress, amsmath,amssymb,prl, aps]{revtex4-1}

\usepackage[T1]{fontenc} 
\usepackage{amsmath, amsthm, amsfonts, amssymb,amstext}
\usepackage{mathptmx}
\usepackage[mathscr]{eucal}
\usepackage{bm}
\usepackage{xfrac}
\usepackage{mathrsfs}
\usepackage{latexsym}
\usepackage{array}
\usepackage{verbatim}
\usepackage{graphicx}
\usepackage{bbold}
\usepackage[normalem]{ulem}
\usepackage[toc,page]{appendix}
\usepackage{color}
\usepackage{tikz}
\usepackage{subfigure}
\usepackage{float}
\usepackage{soul}
\usepackage{mathptmx}
\usepackage{amssymb}
\usepackage{amsmath}
\usepackage{amsfonts}
\usepackage{array}
\usepackage{bm}
\usepackage{xcolor}
\usepackage{braket}
\definecolor{mygrey}{gray}{0.35}
\definecolor{myblue}{rgb}{0.2,0.2,0.8}
\definecolor{myzard}{cmyk}{0,0,0.05,0}
\definecolor{mywhite}{rgb}{1,1,1}
\definecolor{myred}{rgb}{1,0.,0.3}
\usepackage[colorlinks=true,citecolor=myblue,linkcolor=myred]{hyperref}

\def\beq{{\begin{equation}}}
\def\eeq{{\end{equation}}}
\def\k{{\rm k}}

\def\i{{\rm i}}
\def\x{{\rm x}}

 \def\ii{\mathord{\rm i}}

\def\half{\textstyle\frac{1}{2}}

\renewcommand{\ii}{{\rm i}}
\def\beq{\begin{equation}}
\def\eeq{\end{equation}}
\def\barray{\begin{eqnarray}}
\def\earray{\end{eqnarray}}

\DeclareFontEncoding{LS1}{}{}
\DeclareFontSubstitution{LS1}{stix}{m}{n}
\DeclareSymbolFont{symbols4}{LS1}{stixbb}{m}{it}
\DeclareMathSymbol{\varhexagonblack}{\mathord}{symbols4}{"DD}
\DeclareMathSymbol{\hexagonblack} {\mathord}{symbols4}{"DE}

\graphicspath{{./gfx/}}
\usepackage{natbib}
\usepackage{color}  
\usepackage{xcolor}
\definecolor{celeste}{cmyk}{1.00, 0.00, 0.00, 0.4}
\usepackage{hyperref}
\hypersetup{
	colorlinks=true, 
	linktoc=all,   
	breaklinks=true, 
 linkcolor=celeste, 
 citecolor=celeste, 
 urlcolor=celeste
}

\def\beq{{\begin{equation}}}
\def\eeq{{\end{equation}}}
\def\k{{\rm k}}

\def\i{{\rm i}}
\def\x{{\rm x}}

 \def\ii{\mathord{\rm i}}

\def\half{\textstyle\frac{1}{2}}

\renewcommand{\ii}{{\rm i}}
\def\beq{\begin{equation}}
\def\eeq{\end{equation}}
\def\barray{\begin{eqnarray}}
\def\earray{\end{eqnarray}}

\begin{document}
\title{Hybrid quantum-classical simulations of semiclassical gravity}
\author{C. Fulgado-Claudio}
\affiliation{Instituto de F\'isica Te\'orica UAM-CSIC, Universidad Aut\'onoma de Madrid, Cantoblanco, 28049, Madrid, Spain} 
\author{D. Gonz\'alez-Cuadra}
\affiliation{Instituto de F\'isica Te\'orica UAM-CSIC, Universidad Aut\'onoma de Madrid, Cantoblanco, 28049, Madrid, Spain} 
\author{J. Beltr\'an Jim\'enez }
\affiliation{Departamento de F\'isica Fundamental, Universidad de Salamanca, E-37008 Salamanca, Spain.}
\affiliation{
Instituto Universitario de F\'isica Fundamental y Matem\'aticas (IUFFyM), Universidad de Salamanca, E-37008 Salamanca, Spain} 
\author{A. Bermudez}
\affiliation{Instituto de F\'isica Te\'orica UAM-CSIC, Universidad Aut\'onoma de Madrid, Cantoblanco, 28049, Madrid, Spain} 

\begin{abstract}
We propose a hybrid quantum-classical algorithm for the simulation of real-time dynamics in interacting quantum field theories coupled to classical  fields, focusing on the  self-consistent estimation of semiclassical backreaction. By discretizing space and time, we construct an {iterative}  protocol that simulates the {Trotterized dynamics} of the quantum fields subject to the dynamical classical fields. 
By estimating certain quantum expectation values 
through a set of projective measurements, we 
source the equations of motion of the classical fields, and solve them  numerically to  feed them forward to the quantum simulation in an iterative self-consistent loop. 
Semiclassical backreaction is 
relevant in various fields of physics, particularly in 
cosmology, where  quantum matter fluctuations affect the gravitational field dynamics, and a controlled renormalization 
must be carefully considered to get a sensible continuum limit. 
We benchmark our algorithm in this context, focusing on scalar-tensor theories of modified gravity  exhibiting a chameleon mechanism, such that  
 a light classical scalar field driving cosmic acceleration becomes massive  in high-density regions, effectively screening 
any possible yet unobserved fifth force. By focusing on numerically tractable regimes, we explicitly show the convergence and robustness of our algorithm when considering the continuum limit and the effect of quantum shot noise. 
Our work paves the way for future experiments 
exploring other non-tractable regimes, including non-perturbative interactions of the quantum fields and how these can change  backreaction and the gravitational dynamics. 
\end{abstract}

\maketitle
\textit{Introduction.--} Semiclassical quantum field theories (QFTs) dealing with  quantum fields evolving on a  
dynamical classical background occupy a prominent place in contemporary theoretical physics. This type of theories describe phenomena in high-energy physics, including Schwinger pair production \cite{PhysRev.82.664,FEDOTOV20231} or the chiral magnetic effect \cite{Kharzeev_2014,ENDRODI2025104153}; in ultrafast and strong-field physics, such as high-harmonic generation and multi-photon ionization \cite{RevModPhys.72.545,Ciappina_2017}; or  high-harmonic spectroscopy in condensed matter as a new tool to probe  the electronic band structure, topological properties and many-body correlations~\cite{RevModPhys.90.021002,Heide2024}. 
At cosmological scales, the evolution of quantum fluctuations in the classical background of an expanding spacetime and an inflaton field also falls under this semiclassical framework \cite{Birrell:1982ix,Parker:2009uva,ford1997quantumfieldtheorycurved,MUKHANOV1992203}. In the standard model of cosmology, the semiclassical field equations
\beq
\label{eq:einstein}
 G_{\mu\nu}=\frac{8\pi G}{c^4}\big\langle \hat{T}_{\mu\nu}\big\rangle^{\rm ren},
\eeq
take into account the mutual backreaction between the classical gravitational field encoded in  Einstein's tensor $G_{\mu\nu}$, and a collection of interacting quantum fields entering through their respective stress-energy tensor $\hat{T}_{\mu\nu}$.

Across all these domains, classical background fields shape the quantum fluctuations and in some of them, most notably in cosmology, the quantum fluctuations can feed back on the classical background and induce nonlinear effects. Once interactions are present, quantum fields can develop non-Gaussian correlations whose expectation values source non-linear evolution of the background, requiring a self-consistent approach. This form of {\it nonlinear semiclassical dynamics} poses a clear computational challenge: the quantum sector is interacting and entanglement grows in time, while the response of background fields depends on non-Gaussian corrections.

These considerations motivate the idea of using quantum computers as quantum simulators~\cite{feynman1982simulating,cirac2012goals,Altman_2021} for semiclassical field theories. Quantum devices can efficiently encode highly entangled non-Gaussian states of interacting quantum fields, and evolve them accurately in real time~\cite{Bauer_2023, DiMeglio_2024}. In fact, they have already been considered in the context of semiclassical approaches, prominently within QFTs in curved spacetimes \cite{ PhysRevLett.85.4643,PhysRevA.68.053613,PhysRevA.69.033602,Lahav_2010,Weinfurtner_2011,2019NatPh..15..785H,Kolobov_2021,Viermann_2022,Fulgado_Claudio_2023,_van_ara_2024,Fulgado_Claudio_2025}, where the general trend has been to treat the classical background as fixed, neglecting the important yet challenging problem of backreaction~\cite{PhysRevD.13.2720,BROUT1995329}. Notable exceptions have addressed platform-specific backreaction in selected analogue systems, including classical-fluid acoustic systems \cite{Balbinot_2005,balbinot_hawking_2005,Patrick_2021} and Bose-Einstein condensates within number-conserving formulations \cite{Sch_tzhold_2005,Baak_2022,Pal_2024}, where backreaction enters beyond the leading-order fixed-background description.

In this work, we present a step forward in this direction by developing a general hybrid quantum-classical algorithm~\cite{Peruzzo2014,Cerezo2021,PhysRevA.106.010101} for simulating nonlinear semiclassical field theories. We benchmark it with a  scalar--tensor theory as a proxy of gravitational backreaction  beyond Eq.~\eqref{eq:einstein}. 
The key idea is to encode the quantum sector of the theory onto a quantum device, which implements the real-time quantum field dynamics and allows for measuring a relevant set of observables. When focusing on quantum computers, this step should combine qubit encoding of the fields with Trotter methods~\cite{Childs_2021} or quantum signal processing~\cite{Low_2019} to evolve them in time. A classical computer is then used to evolve the background fields by numerically solving the classical equations of motion sourced by expectation values of the measured observables. In this process, some subtleties must be addressed, most notably the renormalization of the expectation values and the convergence to the continuum limit. Our closed hybrid loop captures the essence of nonlinear semiclassical dynamics and can be applied to a broad set of problems, from inflation  to modified gravity, or even to semiclassical regimes of other disciplines, changing the specific classical equations of motion and the discretized Hamiltonian QFT accordingly. 

\textit{Semiclassical theories.--} The set of semiclassical theories we consider is determined by two main ingredients: \textit{(i)} a quantum Hamiltonian, which depends on a set of dynamical classical degrees of freedom, and governs the real-time evolution of the quantum sector of the theory, and \textit{(ii)} a set of equations of motion in the classical sector sourced by expectation values of the quantum fields. Denoting the classical and quantum fields by $\Phi\equiv\{\phi_i(x)\}_{i=1}^n$ and $\hat\Psi\equiv\{\hat{\psi}_i(x)\}_{i=1}^m$, respectively, where $x=(t,\;\boldsymbol{x})$ denotes the spacetime coordinates, a generic semiclassical theory in $D=d+1$ spacetime dimensions reads
\begin{align}
&\frac{\delta S_{\rm cl}}{\delta\phi_i}=-
\sum_{l}\frac{\partial\lambda_l}{\partial\phi_i}\braket{\hat{O}_l[\hat\Psi,\;\hat{\Pi}_\psi]}^{\rm ren},\label{eq:classical_equations}\\&\frac{{\rm d}}{{\rm d}t}\hat{O}_l=-\ii\Big[\hat{H}[\hat\Psi,\;\hat\Pi_\psi],\;\hat{O}_l[\hat\Psi,\;\hat\Pi_\psi]\Big],
\label{eq:quantum_equations}
\end{align}
where $\hat{H}=\int {\rm d}^dx\sum_{l} \lambda_l(\Phi)\hat{O}_l[\hat\Psi,\;\hat{\Pi}_\psi]$ is the operator-valued Hamiltonian functional, and $S_{\rm cl}$ is the action functional of  the classical field theory. Here, $\lambda_l(\Phi)$ are a set of microscopic couplings of the quantum fields and can also depend on the classical fields. $\hat{O}_l[\hat\Psi,\;\hat{\Pi}_\psi]$ may include arbitrary products of the quantum fields, their spatial derivatives, and their conjugate momenta $\hat\Pi_\psi$. In this expression, $\braket{\bullet}^{\rm ren}$ is a properly renormalized expectation value. In the following, we focus on problems in which the  classical fields are homogeneous, $\Phi(x)=\Phi(t)$, noting that the structure can be readily generalized to inhomogeneous cases.

Note that when $\lambda_l=0$ for all operators $\hat{O}_l$ containing products of quantum fields beyond quadratic order, the quantum sector can be solved efficiently using classical Gaussian methods, since all the dynamics is contained in the two-point correlations \cite{Peschel_2003}. However, in standard QFTs of relevance where interactions are the key, e.g. gauge theories with dynamical fermion matter, the computation of non-perturbative real-time dynamics, which can also lead to finite fermion densities, pose severe limitations to our current classical approaches, as one  faces the sign problem or the entanglement barrier 
hindering a fully self-consistent description of backreaction. In this work, we propose an algorithm that overcomes these problems. 

\textit{The algorithm.--} To go beyond the classically-tractable regime, we propose a hybrid quantum-classical algorithm in which both sectors of the theory are encoded using quantum and classical processors, respectively, and propagated in parallel and self-consistently in a controlled iterative manner. To implement the quantum sector of the theory, we consider a simulation of the quantum real-time evolution~\eqref{eq:quantum_equations} on a quantum device. We begin by regularizing the continuum QFT via a Hamiltonian lattice field theory (LFT). This LFT can either be directly amenable to analog quantum simulators by e.g. a truncation into a generalised spin/boson/fermion  models using trapped-ion crystals~\cite{Blatt2012,RevModPhys.93.025001}, arrays of Rydberg atoms \cite{browaeys_many-body_2020,morgado_quantum_2021} or ultracold atoms in optical lattices \cite{Lewenstein01032007,bloch_quantum_2012,halimeh_cold-atom_2025}), or suitable for a qubit encoding on a digital quantum processor \cite{doi:10.1126/science.273.5278.1073,fauseweh_quantum_2024}. The LFT is characterized by a spatial lattice of spacing $a$ and $N_S$ sites, which define an infrared (IR) cutoff $L=aN_S$ (see Supplemental Material in \cite{SM}). The correct approach to the continuum limit, considering renormalizations under varying the ultraviolet (UV) cutoff $a$, is discussed below.

\begin{figure}
   \centering
   \includegraphics[width=1\linewidth]{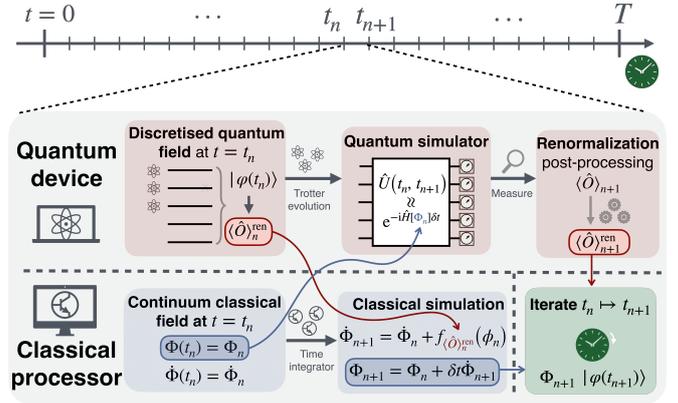}
   \caption{Pictorial representation of the hybrid classical-quantum algorithm. 
   }
   \label{fig:setup_backreaction}
 \end{figure}

Once a specific mapping to  the quantum device is chosen, we approximate the time evolution for both the classical and quantum sectors self-consistently for $t\in[0,\;T]$. We discretize time by defining grid points $\{t_n\}_{n=0}^{N_t}$ with uniform spacing $\delta t\equiv t_{n+1}-t_n$, with $t_0=0$, $t_{N_t}=T$. We denote $\Phi_n\equiv\Phi(t_n)$ and $\braket{\hat{O}_l}_n\equiv\braket{\hat{O_l}(t_n)}$. For the classical fields $\Phi$, which evolve according to the classical equations of motion \eqref{eq:classical_equations}, we employ a numerical integrator such as the symplectic implicit Euler (SIE). This yields an update rule $\Phi_{n+1}=f_{\braket{\hat{O}}_n^{\rm ren}}(\Phi_n)$ depending on certain renormalized expectation values $\braket{\hat{O}}^{\rm ren}_n\equiv\sum_{l}\frac{\partial\lambda_l[\Phi]}{\partial\Phi}\braket{\hat{O}_l[\hat\Psi,\;\hat\Pi_\Psi]}^{\rm ren}_n$ that encapsulate the backreaction of the quantum sector onto the classical fields. For the quantum sector, we use a Trotter-Suzuki approach \cite{Trotter_59,Suzuki_76} to approximate the time evolution in Eq.~\eqref{eq:quantum_equations}, which explicitly depends on $\Phi_n$. Each Trotter step $U(t_n,\; t_{n+1})\approx\exp(-\ii \hat{H}(\Phi_n)\delta t)\equiv U(\Phi_n)$ is then built from a sequence of native quantum gates or analog infinitesimal evolutions, updating the state $|\varphi(t_{n+1})\rangle\approx U(\Phi_n)|\varphi(t_n)\rangle$. At the end of each iteration, we store the classical value of $\Phi_n$ and measure the observables $\braket{\hat{O}_l}_n$ on the quantum device. Then, in order to subtract UV divergences, we apply an appropriate renormalization prescription via classical post-processing, which in curved spacetimes and dynamical scenarios can be implemented via an adiabatic subtraction~\cite{Birrell:1982ix,Landete_2013,Barbero_G__2018} to the required order, thus obtaining $\braket{\hat{O}}_n^{\rm ren}$. These expectation values have a well-defined continuum limit (see Supplemental Material in \cite{SM}), and can then be used to calculate the subsequent Trotter and classical updates, respectively. Iterating this quantum-classical loop forward in time allows to recover the full semiclassical dynamics, as schematically depicted in Fig.~\ref{fig:setup_backreaction}. 

\textit{Chameleon mechanism.--} Screening mechanisms in modified gravity allow to incorporate scalar fields with relevant effects on cosmological scales while evading stringent bounds on fifth forces from local gravity tests \cite{Joyce_2015}. The chameleon mechanism \cite{Khoury_2004_2,Khoury_2004} yields a concrete screening in which the scalar field acquires a density-dependent mass that effectively suppresses the associated fifth forces in our high-density laboratory gravity tests. Chameleon-like models are thus viable extensions beyond the standard cosmological model and, crucially, offer an ideal testbed to benchmark our hybrid quantum–classical algorithm as semiclassical backreaction is instrumental. 
 
To study the chameleon mechanism, we first consider a scalar–tensor theory and simplify the experimental requirements by considering a $D=1+1$ dimensional spacetime. Unlike full General Relativity \eqref{eq:einstein}, where the Einstein tensor vanishes identically in $1+1$ dimensions, this model retains a nontrivial self-consistent gravity-matter backreaction through the scalar field. Considering fermionic fields as matter content, the action of the scalar-tensor theory reads
\begin{align}
S&=\int \!\!{\rm d}^2 x \sqrt{-g}\left[-\frac{M_{\rm Pl}^2}{2}R[g_{\mu\nu}]+\half g^{\mu\nu}\partial_\mu\phi\partial_\nu\phi-V_{\rm cl}[\phi]\right]\nonumber\\&+\int\!\! {\rm d}^2x \sqrt{-\tilde{g}}\left[\overline\Psi\left(\ii\tilde{g}_{\mu\nu}\tilde\gamma^\mu\tilde\nabla^\nu-m\right)\Psi+\frac{g_0^2}{2}(\overline{\Psi}\Psi)^2\right].\label{eq:scalar-tensor action}
\end{align}
Here, $M_{\rm Pl}$ is the Planck mass, $g_{\mu\nu}$ is a metric tensor in the Einstein frame, $R[g_{\mu\nu}]$ is its associated Ricci scalar and $\phi$ is a scalar field responsible for a quintessential gravitational force with potential $V_{\rm cl}[\phi]$ that we will assume to be spatially homogeneous. The matter sector contains a Dirac field $\Psi$ defined on a Jordan frame with metric $\tilde{g}_{\mu\nu}=A^2[\phi]g_{\mu\nu}$, $\tilde\gamma^\mu$ are the curved gamma matrices fulfilling $\{\tilde\gamma^\mu,\;\tilde\gamma^\nu\}=2\tilde g^{\mu\nu}$, and $\tilde\nabla_\mu$ is the covariant derivative for fermions. We also include a fermion self-interaction, connecting to the Gross-Neveu model \cite{PhysRevD.10.3235}, which shares key features with higher-dimensional gauge theories such as dynamical mass generation \cite{RevModPhys.64.649}. By considering a flat metric $g_{\mu\nu}\to\eta_{\mu\nu}$, we can isolate the scalar-fermion interactions. This model is thus determined up to the choice of $V_{\rm cl}[\phi]$ and $A[\phi]$.

Quantizing the matter sector following semiclassical treatments in QFT in curved spacetimes (QFTCS) \cite{Birrell:1982ix,Parker:2009uva,ford1997quantumfieldtheorycurved} yields the two fundamental ingredients announced in Eqs.~\eqref{eq:ST_classical_equation}-\eqref{eq:ST_quantum_equations} (see Supplemental Material in \cite{SM}). First, at the level of the equation of motion for the scalar field, we renormalize the quantum expectation values through a time-dependent normal ordering \cite{Figueroa_2013}, arriving at
\begin{equation}
\frac{\partial^2\phi}{\partial t^2}=-\frac{\delta V_{\rm cl}[\phi]}{\delta \phi}-m\frac{\delta A[\phi]}{\delta\phi}\braket{\hat{\overline\Psi}(x)\hat\Psi(x)}^{\rm ren},\label{eq:ST_classical_equation}
\end{equation}
such that the fermion condensate directly affects the scalar field dynamics. For the fermions, after rescaling the Dirac quantum fields as $\hat\Psi\to A^{-1/2}[\phi]\hat{\Psi}$, we arrive at the Hamiltonian
\begin{align}
\hat{H}=\int \!\!{\rm d}{\rm x}\;\hat{\overline{\Psi}}(x)\left(-\ii\gamma^1\partial_{\rm x}+mA[\phi(t)]\right)\hat{\Psi}(x)-\frac{g_0^2}{2}(\hat{\overline{\Psi}}\hat\Psi)^2.\label{eq:ST_quantum_equations}
\end{align}
 
\begin{figure}[!t]
\centering
\includegraphics[width=\linewidth]{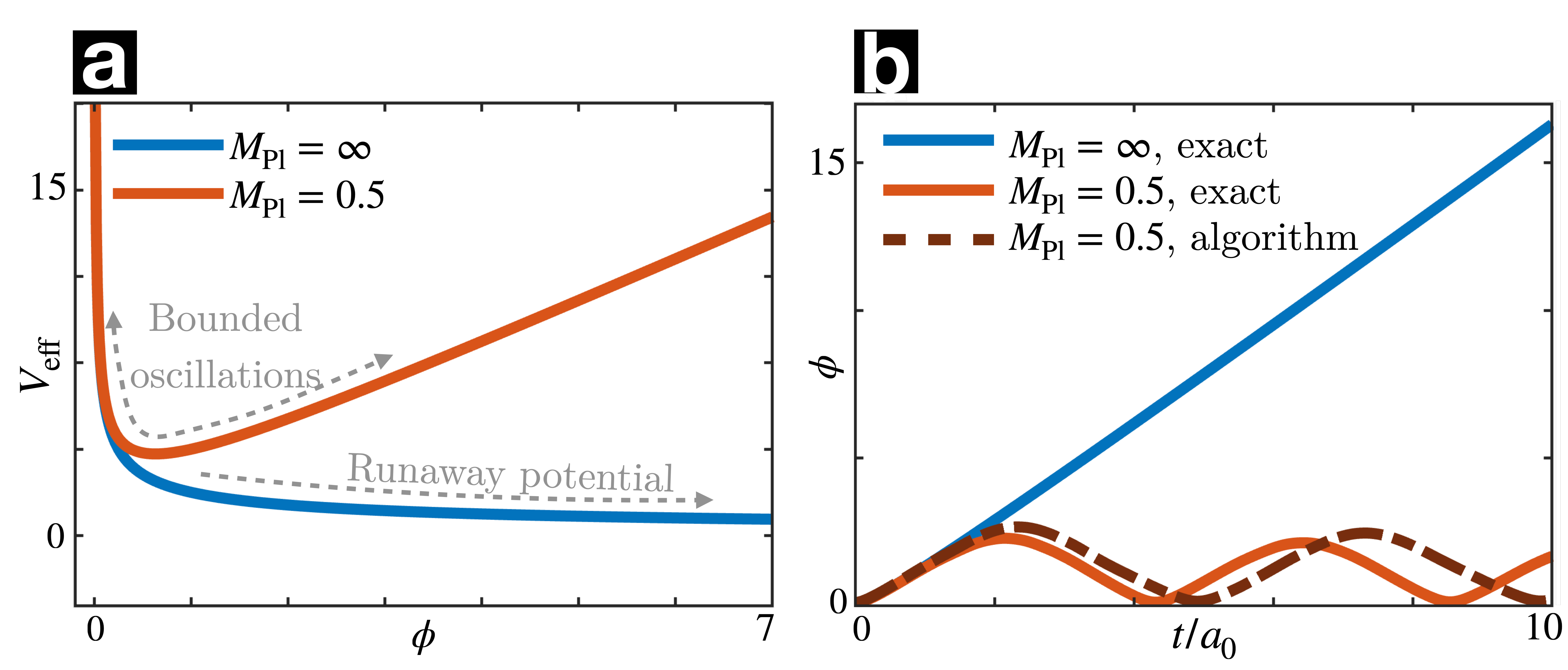}
\caption{{\bf (a)} Representation of the effective potential for the classical scalar field in the absence and presence of backreaction. {\bf (b)} Solutions of the backreaction equations \eqref{eq:ST_classical_equation}-\eqref{eq:ST_quantum_equations} for the chameleon model \eqref{eq:model_functions}, in the absence and presence of backreaction. In the latter case, the exact (continuum) and the algorithm-generated solutions are represented. Parameters are fixed, with respect to a reference energy scale $a_0$, to $ma_0=0.1$, $M_{\rm Pl}=0.5$ (backreaction) or $M_{\rm Pl}=\infty$ (no backreaction), $\lambda a_0^2=0.5$, $\xi=0.5$, $T/a_0=10$, with initial conditions $\phi_0=0.1$ and $\dot\phi_0/a_0=0$. The dashed line shows the real-time evolution of the scalar field obtained by the hybrid algorithm. The algorithm parameters are fixed as $L/a_0=4$, $a/a_0=0.5$, $N_S=8$ and $\tilde{\delta t}=0.1$. 
} 
\label{fig:chamelelon_continuum}
\end{figure}
 
The functionals $V[\phi]$ and $A[\phi]$ in \eqref{eq:scalar-tensor action} are chosen such that the resulting theory displays the aforementioned density-dependent chameleon screening \cite{Khoury_2004_2}. A common realization employs a runaway potential and an exponential conformal factor, that arises in modified gravity and extra-dimension models~\cite{Clifton:2011jh,Joyce:2014kja},
\begin{equation}
V_{\rm cl}[\phi]=\lambda\phi^{-\xi},\;\;\;A[\phi]=\exp(\phi/M_{\rm Pl}).\label{eq:model_functions}
\end{equation} 

For the choice in Eq.~\eqref{eq:model_functions}, the scalar equation \eqref{eq:ST_classical_equation} reduces to a runaway evolution when the renormalized scalar condensate  vanishes, $\langle \hat{\overline\Psi}(x)\hat{\Psi}(x)\rangle^{\rm ren}=0$, with $\phi$ rolling down $V_{\rm cl}(\phi)$. In contrast, when $\langle \hat{\overline\Psi}(x)\hat{\Psi}(x)\rangle^{\rm ren}\neq 0$, the sourcing through $A(\phi)$ generates an effective potential with a density-dependent minimum, endowing $\phi$ with a large effective mass and yielding bounded oscillations about that minimum (see Fig.~\ref{fig:chamelelon_continuum}{\bf (a)}), thus screening the fifth forces at high fermion densities. Since the time evolution of $\phi(x)$ and that of $\braket{\hat{\overline\Psi}(x)\hat\Psi(x)}$ are intertwined, a complete solution of this model relies on a self-consistent treatment of the genuine backreaction. 

The classical-quantum system~\eqref{eq:ST_classical_equation}-\eqref{eq:ST_quantum_equations} falls into the form of Eqs.~\eqref{eq:classical_equations}-\eqref{eq:quantum_equations}, and can thus be implemented using our hybrid quantum-classical algorithm. Furthermore, the model can be solved exactly using Gaussian methods when setting $g_0^2=0$, as the quantum Hamiltonian is quadratic. This provides a natural benchmark for a quantitative assessment of the accuracy of our approach, as we can efficiently account for the required renormalization procedure in real time and the effect of finite measurement resources without actually running the algorithm on quantum hardware. In Fig.~\ref{fig:chamelelon_continuum}{\bf (b)}, we show the exact solution of this chameleon-like semiclassical field theory, obtained by numerically integrating the coupled continuum ordinary differential equations derived from the Dirac equation and from the classical scalar-field action, both with and without backreaction, where the screening mechanism becomes evident. We then compare this result with the time evolution produced by our hybrid algorithm, assuming a staggered fermion discretization of the Hamiltonian \cite{PhysRevD.11.395,PhysRevD.16.3031} with periodic boundary conditions, and employing a SIE integrator for the classical sector (see Supplemental Material in \cite{SM}). 
 
This comparison shows that our algorithm qualitatively reproduces the chameleon mechanism driven by backreaction. The quantitative deviations from the continuum solution follow from {\it (i)} lattice artifacts associated with the spatial discretization of the quantum sector, and {\it (ii)} the finite time step $\delta t$ used to evolve the coupled system. In the next section, we analyze them in more detail and present a systematic strategy to eliminate both sources of error by carefully considering the spacetime continuum limit. 

\begin{figure}
\centering
\includegraphics[width=\linewidth]{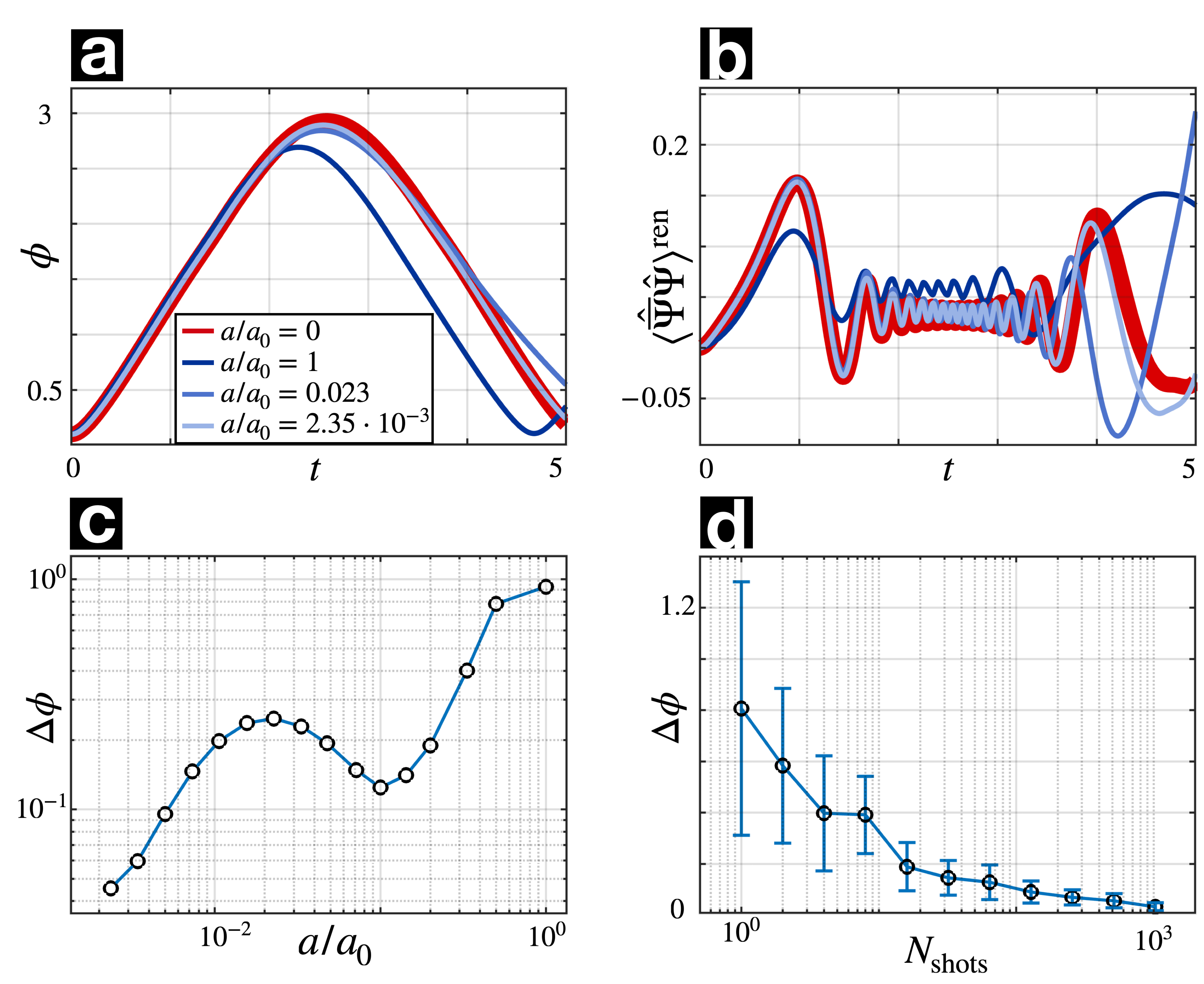}  
\caption{{\bf (a)-(b)} 
Comparison between the continuum solution and the algorithm solutions for the time evolution of $\phi$ and $\braket{\hat{\overline\Psi}\hat\Psi}^{\rm ren}$, respectively, for different values of the lattice spacing $a$. The parameters, in $a_0$ units, are fixed as $ma_0=0.1$, $M_{\rm Pl}=0.5$, $\lambda a_0^2=0.5$, $\xi=0.5$, $T/a_0=5$, with initial conditions for the scalar field $\phi_0=0.1$ and $\dot\phi_0/a_0=0$. The algorithm parameters are $L/a_0=4$, $\tilde{\delta t}=0.05$. {\bf (c)} $\mathrm{L}^2$ distance between the algorithm-generated time evolution of the classical scalar field and the continuum solution, with matching IR cutoff. The parameters are the same as in {\bf (a)-(b)}. 
{\bf (d)} $\mathrm{L}^2$ distance between the algorithm-generated time evolution of the classical scalar field in presence and absence of shot noise as a function of the number of shots $N_{\rm shots}$. The plot points are given by the mean of the $\mathrm{L}^2$ distance for $N_{\rm rep}=20$ repetitions, and the error bars by their standard deviation. The parameters are fixed, with respect to a reference energy scale $a_0$, to $ma_0=0.1$, $M_{\rm Pl}=0.5$, $\lambda a_0^2=0.5$, $\xi=0.5$, $T/a_0=10$, with initial conditions for the scalar field $\phi_0=0.5$ and $\dot\phi_0/a_0=0$, and algorithm parameters $L/a_0=32$, $a/a_0=0.5$, $\tilde{\delta t}=0.05$. 
} 
\label{fig:cont_limit}
\end{figure}

\textit{Continuum limit and renormalization.--} 
So far, we have considered a quantum device to simulate the real-time evolution of a LFT that regularizes Eq.~\eqref{eq:scalar-tensor action}, but the final goal is to address problems in the continuum $a\to 0$. 

Being the scalar field homogeneous, the classical equations of motion~\eqref{eq:ST_classical_equation} do not rely on this discretization, and the continuum limit only involves the quantum sector, requiring a renormalization of the observables that act as sources $\braket{\hat{O}_l[\hat\Psi,\;\hat\Pi_\psi]}\to\braket{\hat{O}_l[\hat\Psi,\;\hat\Pi_\psi]}^{\rm ren}$. Additionally, since the method relies on approximations to the exact time evolution, we must additionally perform a $\delta t\to0$ limit. These two processes can be performed simultaneously by defining the time grid spacing in lattice units, $\tilde{\delta t}=\delta t /a$, and keeping its value fixed. 
 
In general, $a$ is not an explicit parameter of the quantum device, and it cannot be directly tuned to the $a\to0$ limit. To adjust it, following \cite{Carena_2021, Bauer_25}, we must instead tune the bare parameters of the Hamiltonian expressed as dimensionless quantities, i.e. in lattice units. For this, bare parameters are first expressed with respect to a reference lattice spacing $a_0$, and their dimensionless values are subsequently adjusted such that the resulting Hamiltonian describes a theory with a rescaled lattice spacing $a=a_0/s$ with $s>1$. This process is iterated until convergence. For quadratic Hamiltonians this tuning follows directly from a mass-dimension scaling analysis, whereas for interacting theories one must instead follow the renormalization group (RG) \cite{Shankar_1994} flows by extracting the corresponding $\beta$-functions. These can be obtained by comparing lattice-computed dimensionless quantities and their measured, dimensionful counterparts, as outlined in \cite{Carena_2021,Bauer_25}, thus determining how the effective couplings flow as the continuum limit is approached. Additionally, to remove IR effects, one must also take the limit $L\to\infty$, which amounts to an increase of the hardware resources.

Figures~\ref{fig:cont_limit}{\bf (a)-(b)} show the convergence towards the $a\to0$ limit of the scalar field and the sourcing scalar condensate $\braket{\hat{\overline\Psi}\hat\Psi}^{\rm ren}$, respectively. The convergence is more quantitatively studied for a wider range of lattice spacings $a$ in Fig.~\ref{fig:cont_limit}{\bf (c)}, where we compute the $\mathrm{L}^2$ distance $\Delta\phi $ between the scalar field resulting from our quantum-classical algorithm and the exact continuum solution with a fixed infrared cutoff matching that of our algorithm (see Supplemental Material in \cite{SM}). We find a non-monotonous dependence for large $a$, which can be traced back to  lattice effects. As soon as the cutoff $\Lambda=\pi/2a$ surpasses the relevant scale in which dynamics take place, the algorithm evolution monotonically approaches the continuum limit as $a\to0$. 

{\it Experimental considerations.--} Approaching the continuum limit requires increasing resources as the lattice spacing $a$ is reduced. After selecting a fermion encoding, depending on factors such as the boundary conditions or the dimensionality of the LFT \cite{Zohar_2019, Derby_2021, Maskara_2025}, the real-time evolution is implemented through a Trotter approach using a sequence of native interactions in the analog platform  or a sequence of quantum gates in the digital one~\cite{10.1063/1.5088164,morgado_quantum_2021}. Keeping the rescaled time step $\tilde{\delta t}=\delta t/a$ fixed requires that, as the lattice spacing is reduced from $a_0$ to $a$, the number of Trotter steps increases as $N_t\propto a_0/a$, leading to increasingly deep circuits. Likewise, maintaining a fixed physical system size $L=aN_S$ requires the number of lattice sites, and thus the number of qubits, to scale as $N_S\propto a_0/a$. These combined constraints ultimately limit how close one will approach the continuum limit on a quantum device in practice. 

We further account for shot noise  in the quantum measurement by treating the outcome of measuring an operator $\hat{O}$ as a random variable with mean $\braket{\hat{O}}$. Repeating the measurement $N_{\rm shots}$ times reduces the variance as $1/N_{\rm shots}$. In our fully dynamical setting, where the measurements outcome backreact on the classical field evolution, this protocol enables the reconstruction of the noiseless trajectories, both in the quantum and classical sectors. In Figure~\ref{fig:cont_limit}{\bf (d)}, we have computed $N_{\rm rep}=20$ simulations of the experiment, including the random shot noise, for different values of $N_{\rm shots}$. For each of the repetitions, we have computed the $\mathrm{L}^2$ distance to the noiseless evolution of $\phi(t)$. Then, for each value of $N_{\rm shots}$, we have plotted the mean of the $N_{\rm rep}$ repetitions, with error bars estimated by their standard deviation. The figure demonstrates that the correct backreacting dynamics is recovered for sufficiently large $N_{\rm shots}$, even in the presence of shot noise, with both the distance to the continuum solution and the statistical deviation going to zero as $N_{\rm shots}$ is increased.

\textit{Conclusions and outlook.--} We have introduced a hybrid quantum-classical protocol to simulate semiclassical systems with dynamical backreaction. By discretizing the quantum sector and coupling it to classical degrees of freedom on a discrete grid of time points, we have shown that the protocol faithfully captures physical phenomena that crucially rely on a consistent implementation of backreaction. Importantly, we have demonstrated that the spatial and temporal continuum limits can be taken simultaneously, ensuring convergence to the correct continuum dynamics.

As a concrete demonstration, we have shown that our protocol accurately describes the physics of the chameleon mechanism in a scalar--tensor theory of gravity, whose screening behavior relies fundamentally on the interplay between quantum fluctuations and classical dynamics. Finally, we have shown that performing the continuum limit in conjunction with averaging over repeated measurements recovers the correct continuum time evolution of the coupled systems. This demonstrates that the protocol is robust against shot noise and that it converges reliably to the expected continuum semiclassical dynamics in the presence of backreaction. 

A natural next step is to apply this hybrid framework to other semiclassical theories that have so far remained inaccessible to classical numerical methods, such as scalar--tensor gravity theories with fully interacting quantum matter or scenarios involving nonperturbative, non-Gaussian quantum fluctuations. In the long term, extensions of this approach to higher-dimensional settings may enable the investigation of semiclassical regimes relevant to general relativity, contingent on the ability to simulate quantum matter fields in $3+1$ dimensions. 

\begin{acknowledgements}
\textit{Acknowledgments.--} C.F.-C. and A.B. acknowledge support from the European
Unions Horizon Europe research and innovation programme
under grant agreement No 101114305 (MILLENION-SGA1 EU Project), from 
PID2024-161474NB-I00 (MCIU/AEI/FEDER,UE), and from
QUITEMAD-CM TEC-2024/COM-84. We also acknowledge support from the Grant IFT Centro de Excelencia Severo Ochoa CEX2020-001007-S funded by
MCIN/AEI/10.13039/501100011033, and from the CSIC Research Platform on Quantum Technologies PTI-001. D.G.-C. acknowledges financial support through the Ramón y Cajal Program (RYC2023-044201-I), financed by MICIU/AEI/10.13039/501100011033 and by the FSE+. JBJ acknowledges support from grants PID2024-158938NB-I00 funded by MICIU/AEI/10.13039/501100011033 and by “ERDF A way of making Europe” and the Project SA097P24 funded by Junta de Castilla y Le\'on.
\end{acknowledgements}

\bibliographystyle{apsrev4-1}
\bibliography{main.bib}

\clearpage
\onecolumngrid    

\setcounter{page}{1} 

\begin{center}

{\large\bfseries Supplemental Material for ``Hybrid quantum-classical simulations of semiclassical gravity''}

\vspace{5mm}

C.~Fulgado-Claudio,$^{1}$
D.~Gonz\'alez-Cuadra,$^{1}$
J.~Beltr\'an Jim\'enez,$^{2,3}$
A.~Bermudez$^{1}$

\vspace{3mm}

{\small \it
$^{1}$Instituto de F\'isica Te\'orica UAM-CSIC, Universidad Aut\'onoma de Madrid, Cantoblanco, 28049 Madrid, Spain\\
$^{2}$Departamento de F\'isica Fundamental, Universidad de Salamanca, E-37008 Salamanca, Spain\\
$^{3}$Instituto Universitario de F\'isica Fundamental y Matem\'aticas (IUFFyM),\\Universidad de Salamanca, Spain
}

\end{center}

\vspace{0.5cm}
\setcounter{equation}{0}
\renewcommand{\theequation}{S\arabic{equation}}

In this Supplemental Material, we provide details that support the quantum-classical algorithm outlined in the main text. Specifically, we give derivations for the semiclassical equations $(6)-(7)$ of the main text; we derive the discretization scheme employed to implement the Dirac fermionic field in a quantum device; we provide details on the renormalization of the operators feeding the classical equation of motion in the hybrid loop; and we detail some numerical aspects used throughout the main text. 

\section{Semiclassical equations for scalar-tensor theories}
In this section, we derive the main equations for scalar-tensor theories, assuming that we fix the metric to Minkowski and that the matter sector consists of a Dirac field in $D=1+1$ dimensions.  We begin from the most general action describing a scalar field mediating a quintessence-like force
\begin{align}
S=\int {\rm d}^D x \sqrt{-g}\left(-\frac{M_{\rm Pl}^2}{2}R+\half g^{\mu\nu}\partial_\mu\phi\partial_\nu\phi-V_{\rm cl}[\phi]\right)+\int {\rm d}^Dx \sqrt{-\tilde{g}}\mathcal{L}_{\rm m}[\Psi,\;\tilde{g}_{\mu\nu}].
\end{align}
Here, $g_{\mu\nu}$ is a metric tensor in the Einstein's frame, $g=\det(g_{\mu\nu})$, $M_{\rm Pl}$ is the Planck mass, $R$ is the Ricci scalar, $\phi(x)$ is a homogeneous scalar field, $V_{\rm cl}[\phi]$ is its classical potential density, and $\mathcal{L}_m[\Psi,\;\tilde{g}_{\mu\nu}]$ is a matter Lagrangian living in a Jordan frame with metric $\tilde{g}_{\mu\nu}=A^2[\phi]g_{\mu\nu}$. We consider  the matter Lagrangian to consist of Dirac fermions
\beq
\mathcal{L}_m[\Psi,\;\tilde{g}_{\mu\nu}]=\overline{\Psi}(\i\tilde{g}^{\mu\nu}\tilde\gamma_\mu\nabla_\nu-m)\Psi,
\eeq
where $\overline{\Psi}=\Psi^\dagger\gamma^0$ is the adjoint Dirac spinor field, $\tilde\gamma^\mu$ are curved spacetime gamma matrices in the Jordan frame fulfilling a generalization of the Clifford algebra $\{\tilde\gamma^\mu,\;\tilde\gamma^\nu\}=2\tilde g^{\mu\nu}$ and $\nabla_\mu$ is the covariant derivative, defined as $\nabla_\mu\Psi=\partial_\mu\Psi+\Omega_\mu\Psi$, where $\Omega_\mu$ is the so-called spin-connection.

Setting the spatiotemporal dimension to $D=1+1$ and fixing the background metric to Minkowski, $g_{\mu\nu}\to\eta_{\mu\nu}$, the action simplifies due to $R\to0$ and $\tilde{g}_{\mu\nu}\to A^2[\phi]\eta_{\mu\nu}$, yielding
\beq
S=\int {\rm d}^2 x \Big(\half \eta^{\mu\nu}\partial_\mu\phi\partial_\nu\phi-V_{\rm cl}[\phi]+A^2[\phi]\overline{\Psi}\big(\i \tilde{g}^{\mu\nu}\tilde\gamma_\mu\nabla_\nu-m\big)\Psi\Big),
\eeq

This action describes a scalar field living in Minkowski and a Dirac field living in a dynamical, conformal metric with the conformal factor depending on the value of the scalar field. Upon computing the spin connection and the curved-spacetime gamma matrices, and after performing a rescaling of the Dirac field $\Psi\rightarrow A^{-1/2}[\phi]\Psi$, the dynamics of the matter sector is encoded in a dynamical, multiplicative renormalization of the Dirac mass, namely 
\beq
S=\int {\rm d}^2 x \Big(\half \eta^{\mu\nu}\partial_\mu\phi\partial_\nu\phi-V_{\rm cl}[\phi]+\overline{\Psi}\big(\i \eta^{\mu\nu}\gamma_\mu\nabla_\nu-mA[\phi]\big)\Psi\Big),
\label{eq:classical_action}
\eeq
where $\gamma^\mu$ are the conventional gamma matrices $\{\gamma^\mu\;\gamma^\nu\}=2\eta^{\mu\nu}$. After these steps, the resulting action corresponds to that of a scalar and a Dirac field, both living in flat spacetime and interacting through the following coupling
\beq
S_{\rm int}=\int {\rm d}^2 x\;mA[\phi]\overline{\Psi}\Psi.
\eeq

At this stage, and taking into account that $\phi(x)$ will not be promoted to a quantum field, we can apply a variational principle to obtain its equation of motion. Taking into account that we consider a homogeneous scalar field $\phi(x)=\phi(t)$, we arrive at
\beq
    \frac{\partial^2\phi}{\partial t^2}=-\frac{\delta V_{\rm cl}[\phi]}{\delta\phi}-m\frac{\delta A[\phi]}{\delta\phi}\overline{\Psi}\Psi,
    \label{eq:classical_eq_phi}
\eeq
where everything is still classical at this point.

The next step in obtaining the semiclassical equations is to   quantize  the matter fields, which in this case is the Dirac fermions. To this end, the Dirac field and its adjoint must be upgraded to quantum operators, and the Dirac brackets that define the classical theory are upgraded to equal-time anti-commutators, arriving at $\{\hat\Psi_\alpha(t,{\rm x}),\;\hat{\Psi}^\dagger_\beta(t,{\rm y})\}=\delta_{\alpha\beta}\delta({\rm x-y})$. This has two main consequences. First, the upgrade of $\hat\Psi$ from classical to quantum fields requires to perform a modification of the scalar field classical equation \eqref{eq:classical_eq_phi}, since the right-hand side would otherwise contain operator-valued terms. The prescription we follow is to take a properly renormalized expectation value, yielding
\beq
    \frac{\partial^2\phi}{\partial t^2}=-\frac{\delta V_{\rm cl}[\phi]}{\delta\phi}-m\frac{\delta A[\phi]}{\delta \phi}\braket{\overline{\Psi}\Psi}^{\rm ren},
    \label{eq:scalar_equation}
\eeq
which corresponds to Eq.~$(6)$ of the main text. 

On the other hand, the action in Eq.~\eqref{eq:classical_action} displays some classical-valued and some operator-valued terms. From the perspective of the quantum theory of matter, we drop the purely classical terms. After performing a Legendre transformation, one arrives at the quantum Hamiltonian of the Dirac fermions
\beq
\hat{H}_{\rm matter}=\int {\rm d}{\rm x}\,\hat{\Psi}^\dagger\big(-\i\gamma^0\gamma^1\partial_{\rm x}+mA[\phi]\gamma^0\big)\hat\Psi,
\label{eq:matter_Hamiltonian_continuum}
\eeq
which corresponds to Eq.~$(7)$ in the main text.

\section{Lattice discretization}

In this section, we discuss the procedure to discretize the Hamiltonian \eqref{eq:matter_Hamiltonian_continuum}, hence making it amenable to a quantum device. First, we replace the continuum space (since we are in $D=1+1$, it is just a line, $x\in\mathbb{R}$) by a grid of $N_S$ points separated by a distance $a$, constituting a lattice of size $L=aN_S$.
\begin{align}
    x\to na;\;\;\;\int dx\to a\sum_n;\;\;\;\hat\Psi(t,\;{\rm x})\to a^{-1/2}\hat\Psi_n(t);\;\;\;\partial_x\hat\Psi\to\frac{\hat\Psi_{n+1}-\hat\Psi_{n-1}}{2a}.
\end{align}

The coefficient $a^{-1/2}$ in the definition of the lattice fermionic operators makes them dimensionless. Introducing these modifications in the continuum matter Hamiltonian \eqref{eq:matter_Hamiltonian_continuum}, we arrive at
\beq
\hat{H}_{\rm matter}^{\rm latt}=
\sum_n\hat\Psi^\dagger_n\left(-\frac{\i}{2a}\gamma^0\gamma^1\left(\hat\Psi_{n+1}-\hat\Psi_{n-1}\right)-mA[\phi]\gamma^0\hat\Psi_n\right).
\eeq

This lattice Hamiltonian, however, suffers from the so-called fermion doubling \cite{PhysRevD.16.3031}. Different solutions have been proposed to overcome this issue, and we follow the prescription known as staggered discretization \cite{PhysRevD.11.395}, where each internal component of the fermionic field is delocalized in different spatial locations. We apply it by using the following rule
\begin{align}
    &\hat\Psi_n\to\begin{pmatrix}
        \hat\chi_n\\0 
    \end{pmatrix}\text{   if $n$ is even}\\&\hat\Psi_n\to\begin{pmatrix}
        0\\\hat\chi_n 
    \end{pmatrix}\text{   if $n$ is odd}
\end{align}

By doing this, one arrives at the lattice Hamiltonian
\beq
\hat{H}_{\rm matter}^{\rm latt}=\sum_n\left(-\frac{\i}{2a}\left(\hat\Psi_n^\dagger\hat\Psi_{n+1}-\hat\Psi_n^\dagger\hat{\Psi}_{n-1}\right)+(-1)^nmA[\phi]\hat\Psi_n^\dagger\hat\Psi_n\right),
\label{eq:lattice_Hamiltonian}
\eeq

The inhomogeneous mass term generates a folding in the Brillouin zone that excludes the fermion doublers. This forces us to work in a reduced Brillouin zone defined as ${\rm RBZ}=\left\{-\frac{\pi}{2a}+\frac{2\pi j}{aN_S}\right\}_{j=1}^{N_S}$. This is the Hamiltonian to be simulated in a quantum device when performing the hybrid algorithm to solve the backreaction problem. 

\section{Renormalization of quantum operators in dynamical backgrounds} 
In this section, we provide details on how to perform the renormalization step of the quantum operator entering the hybrid loop. As explained in the main text, the ultimate goal of the hybrid algorithm is to accurately describe the continuum backreacting coupled evolution. Since the matter  Hamiltonian implemented on the quantum device is discretized, this requires sending the lattice spacing to 0. In lattice approximations of continuum quantum field theories (QFTs), the lattice spacing leads to two main deviations. Given the Hamiltonian of a continuum field theory $\hat{H}_{\rm QFT}$, the lattice field theory that approximates it, $\hat{H}_{\rm LFT}$ is chosen so that
\beq
\hat{H}_{\rm QFT}=\hat{H}_{\rm LFT}+\mathcal{O}(a^p),\;\;\;p\geq1
\label{eq:QFT-LFT_connection}
\eeq
i.e. both Hamiltonians coincide up to terms that vanish as $a\to0$. This induces a first error source for finite $a$. The second effect of approximating a QFT by a LFT is the introduction of a physical lattice spacing $a$ and a lattice box size $L=aN_S$. In the case of staggered fermions, this restricts the allowed values of the lattice quasimomentum to the reduced Brillouin zone, as explained in the previous section. The distance between consecutive values is $\Delta{\rm k}={\rm k}_j-{\rm k}_{j-1}=\frac{\pi}{L}$, so the size of the box that contains the LFT determines an IR cutoff. On the other hand, the maximum allowed value of the quasimomentum is $|{\rm k}_{\rm max}|=\frac{\pi}{2a}$, identifying the lattice spacing with a UV cutoff. As a consequence of these two results, it is clear that obtaining a good approximation to the continuum QFT relies, by virtue of Eq.~\eqref{eq:QFT-LFT_connection}, on sending $a\to0$, which in turn amounts to sending the UV cutoff to infinity. However, it is a well known result in QFTs that, in general, doing this leads to spurious results due to the presence of UV divergences, which prevents a direct computation of finite observables from the QFT. The procedure to cure these divergences is given by the machinery of the renormalization group, a set of techniques that allow to render the observables of the QFT finite by consecutive coarse-graining and rescaling steps. In this work, however, being the Hamiltonian quadratic, the RG machinery is drastically simplified, and the only renormalization step we must care about is that of a vacuum subtraction, tantamount to a normal ordering in flat spacetimes. 

The observable we are interested in is $\braket{\hat{\overline\Psi}\hat\Psi}$ in a theory of free fermions in $D=1+1$ with a dynamical mass given by $mA[\phi]$, where $A[\phi]$ is self-consistently determined within the hybrid loop. Therefore, we need to renormalize this operator step by step, at the same time as we determine dynamically the real-time evolution of $\phi(t)$. To do so, we follow the time-dependent normal ordering prescription, which amounts to a zeroth-order adiabatic subtraction \cite{Landete_2013,Barbero_G__2018}. 

The Dirac field can be expanded in terms of mode functions and particle/antiparticle operators, 
\beq
\hat\Psi(t,\;{\rm x})=\int\frac{d{\rm k}}{2\pi}\left(u_\k(t)\hat{a}_\k+v_{-\k}(t)\hat{b}^\dagger_{-\k}\right)e^{\i\k\x}.
\eeq

In flat spacetimes with no time-dependence of the parameters, the mode functions $u_\k(t)$ and $v_k(t)$ are naturally determined as plane waves corresponding to positive- and negative-frequency components of the field. However, in dynamical setups, there is an ambiguity in how to determine the notion of particle, and therefore in how to choose the mode functions. Here, we stick to an instantaneous vacuum convention, corresponding to a zeroth-order adiabatic vacuum, which is enough to remove the UV divergences that arises in our model for the $\braket{\hat{\overline\Psi}\Psi}$ operator.

The mode functions $u_{\k}(t)$ and $v_{-\k}(t)$ satisfy the dynamical equations
\begin{align}
    &\i\frac{d}{dt}u_\k(t)=h^{\rm SP}_ku_k(t),\;\;u_\k(0)={\rm v}_\k^+(0),\label{eq:mode_function_u}\\&\i\frac{d}{dt}v_{-\k}(t)=h^{\rm SP}_kv_{-k}(t),\;\;v_{-\k}(0)={\rm v}_\k^-(0),\label{eq:mode_function_v} 
\end{align}
where $h^{\rm SP}_\k$ is the single-particle Hamiltonian defined by $\hat{H}=\int\frac{d\k}{2\pi}\hat\Psi_\k^\dagger h^{\rm SP}_\k\hat\Psi_k$, where $\hat{\Psi}_\k=\int{d\x}\;\hat\Psi(x) {\rm e}^{\i\k\x}$ and ${\rm v}_\k^{\pm}(t)$ are the instantaneous positive- and negative-energy eigenvectors of the matrix $h^{\rm SP}_\k$. In this work, the Hamiltonian corresponds to that in Eq.~\eqref{eq:matter_Hamiltonian_continuum}, so $h_\k^{\rm SP}(t)=-\k\gamma^0\gamma^1+mA[\phi]\gamma^0$. By virtue of the dependence of the mass term on $\phi(t)$, the Hamiltonian is time-dependent, which makes all of the above derivation necessary.

Within the instantaneous vacuum convention, the notion of particle is obtained, for each time $t=t_\star$, by expanding the Dirac field operator in the set of instantaneous eigenvectors of the single-particle Hamiltonian $\rm v_\k^{\pm}(t_\star)$,
\beq
\hat\Psi(t_\star,\;\x)=\int\frac{d\k}{2\pi}\left({\rm v_k^+(t_\star)}\hat{a}_\k^\star+{\rm v_k^-}(t_\star)\hat{b}_{-\k}^{\star\dagger}\right)e^{\rm ikx}
\eeq
The sets of particle and antiparticle operators, $\hat{a}_\k^\star$ and $\hat{b}_{-\k}^{\star}$, define the instantaneous vacuum state $\ket{0_\star}$ by the property $\hat{a}_\k^\star\ket{0_\star}=\hat{b}_{-\k}^{\star}\ket{0_\star}=0$ $\forall\k$. The renormalization scheme that we follow, time-dependent normal ordering \cite{Figueroa_2013}, amounts to subtracting, at each time, the vacuum contribution to the expectation value, so that the resulting  object is UV-convergent. 

Fixing an initial groundstate associated to the time $t=0$, and working within the Heisenberg picture, we can easily derive expressions both for the non-renormalized and for the renormalized expectation vacuum of the bilinear operator in terms of the aforementioned mode functions and instantaneous eigenvectors of the single-particle Hamiltonian.
\begin{align}
    &\braket{\hat{\overline\Psi}(t)\hat\Psi(t)}=\bra{0_0}\hat{\overline\Psi}(t)\hat\Psi(t)\ket{0_0}=\int\frac{d\k}{2\pi}v_{-k}^\dagger(t)\gamma^0v_{-\k}(t),\label{eq:bare_scalar_condensate}\\
    &\braket{\hat{\overline\Psi}(t)\hat\Psi(t)}^{\rm inst}=\bra{0_t}\hat{\overline\Psi}(t)\hat\Psi(t)\ket{0_t}=\int\frac{d\k}{2\pi}{\rm (v_k^-)^\dagger(t)\gamma^0v_k^-(t)},\label{eq:instantaneous_scalar_condensate}\\&\braket{\hat{\overline\Psi}(t)\hat\Psi(t)}^{\rm ren}=\braket{\hat{\overline\Psi}(t)\hat\Psi(t)}-\braket{\hat{\overline\Psi}(t)\hat\Psi(t)}^{\rm inst},\label{eq:renormalized_scalar_condensate}
\end{align}
where $\ket{0_0}$ and $\ket{0_t}$ are the instantaneous vacuum states corresponding to $t=0$ and to a generic time $t$. The interpretation of this subtraction is tantamount to that of normal ordering in traditional Dirac field in Minkowski spacetime: to remove vacuum contributions to the expectation values, hence removing UV-divergent contribution. This computation yields the same result as a zeroth-order adiabatic subtraction \cite{Landete_2013,Barbero_G__2018}. In $D=1+1$ dimensions, this subtraction removes the UV-divergent vacuum contribution at each time, so that the renormalized expectation value has a well-defined continuum limit, so that one can faithfully take the $a\to0$ limit, allowing the hybrid algorithm outlined in the main text to approach the continuum Hamiltonian, $\hat{H}_{\rm LFT}\to\hat{H}_{\rm QFT}$. Higher orders of adiabatic subtraction might be necessary for different theories, such as higher-dimensional Dirac fields.

In the lattice implementation, two changes must be incorporated: {\textit (i)} the Hamiltonian used to obtain the single-particle Hamiltonian, its instantaneous eigenvectors ${\rm v}_\k^\pm(t)$ and to evolve the mode functions $u_\k(t)$ and $v_{-\k}(t)$ is that in \eqref{eq:lattice_Hamiltonian} instead of the one in \eqref{eq:matter_Hamiltonian_continuum}, so that the lattice single-particle Hamiltonian reads $h^{\rm SP,\;latt}_{k}=-\frac{\sin(\k a)}{a}\gamma^0\gamma^1+mA[\phi]\gamma^0$; and \textit{(ii)} the momentum integral is replaced by a sum over the discrete quasimomenta ${\rm k}_j$, so that
\beq
\braket{\overline\Psi(t)\Psi(t)}^{\rm ren}= \frac{1}{L}\sum_{j=1}^{N_S}
\left[v_{-{\rm k}_j}^\dagger(t)\gamma^0 v_{-{\rm k}_j}(t)
-\big({\rm v}_{{\rm k}_j}^-(t)\big)^\dagger\gamma^0{\rm v}_{{\rm k}_j}^-(t)\right],
\eeq
where the single-particle Hamiltonian is replaced by the lattice Hamiltonian with $\mathcal{O}(a^p)$, $p\geq1$ deviations from the continuum one. This is the quantity fed into the classical update rule at each time step.

\section{Numerical details}
In this section, we provide details on different numerical details that are used in the results of the main text. 

First, the hybrid loop requires two numerical ingredients to simulate the time evolution of the classical and quantum sectors: one for the classical and one for the quantum sector. The structure of the algorithm, as represented in Fig.~$1$ of the main text, imposes a restriction on the numerical integrator that we can use in the classical sector. Typically, for a differential equation of the form
\beq
\frac{d}{dt}{\rm f}(t)=F[{\rm f(t)},\;t],
\eeq
a numerical integrator relies on the definition of a time grid with elements $t_n=n\delta t$, where $n\in\mathbb{N}$, and it approximates the full function ${\rm f}(t)$ by a set $\{{\rm f}_n\equiv{\rm f}(t_n)\}_{n=0}^{N_t}$. To obtain the element ${\rm f}_n$, a generic numerical integrator might need to evaluate the functional $F$ for time elements that can be: {\it (i)} in the future, i.e. $F\left[\cdot,\;t_{m}\right]$ with $m>n$, or {\it (ii)} outside the time grid, e.g. $F\left[\cdot,\;t_{n+\half}\right]$. Both of these evaluations are unproblematic in a general scenario, since the dependence of $F$ on $t$ is explicit and given by the differential equation, so it can be evaluated in arbitrary time points. However, our algorithm is constrained by the fact that the differential equation governing the time evolution of $\phi(t)$ is not known {\it a priori}, but rather must be self-consistently determined as one solves the time-evolution of the operator $\braket{\hat{\overline\Psi}\hat\Psi}^{\rm ren}$. This prevents one from evaluating the right hand side of Eq.~\eqref{eq:scalar_equation} for future time steps. Additionally, the time is discretized on the same grid for both the classical and the quantum sectors, so intermediate times are not accessible either. Given these restrictions, in the main text we use a symplectic Euler method, which we found to work well for the problem at hand. It is given by splitting the second-order differential equation for $\phi(t)$ into two first- order differential equations for $\phi$ and $\dot\phi$, and evolving them as
\begin{align}
    &\dot\phi_{n+1}=\dot\phi_n+\delta t\left(-\frac{\delta V_{\rm cl}[\phi]}{\delta\phi}\Bigg|_{\phi=\phi_n}-m\frac{\delta A[\phi]}{\delta\phi}\Bigg|_{\phi=\phi_n}\braket{\hat{\overline{\Psi}}(t_n)\hat\Psi(t_n)}^{\rm ren}\right),\\
    &\phi_{n+1}=\phi_n+\delta t\dot\phi_{n+1}.
\end{align}

Additionally, we also wanted to define the $\mathrm{L}^2$ distance as used in the paper. We use this distance to evaluate how close two different time-evolutions are. Suppose we have two functions of time, ${\rm f}(t)$ and ${\rm g}(t)$ for $t\in[0,\;T]$. We define their $\mathrm{L}^2$ distance as 
\begin{equation}
    d({\rm f},\;{\rm g})=\left(\int_0^T|{\rm f}(t)-{\rm g}(t)|^2dt\right)^{\half}.
\end{equation}

Since the algorithm does not produce functions but sets of values of the functions evaluated at discrete time steps, this distance is generalized to 
\beq
 d({\rm f},\;{\rm g})=\left(\sum_{n=0}^{N_t}|{\rm f}_n-{\rm g}_n|^2\delta t\right)^{\half}.
 \label{eq:L2_distance_discrete}
\eeq

Now, we address the way in which we evaluate the convergence to the continuum limit. In the main text, we state that we compute the $\mathrm{L}^2$ distance between the hybrid algorithm-generated time-evolution of $\phi(t)$ and the continuum-limit solution and computed using a IR cutoff matching that of the hybrid algorithm. By this, we mean the following. First, we obtain the time-evolution of $\phi(t)$ by applying the hybrid algorithm to the semiclassical theory for parameters corresponding to decreasingly large values of $a$, thus obtaining a family of sets that we denote here $\{\phi^a_n\}_{n=0}^{N_t}$, where the sub-index $n$ corresponds to the time step and the super-index $a$ to the corresponding value of the lattice spacing. This is done keeping the value of the lattice size $L$ fixed, which corresponds to fixing the IR cutoff $\Delta k=\frac{\pi}{L}$. Then, to assess its convergence to the $a\to0$ limit, we compute the limiting solution by solving simultaneously the equation for the scalar field \eqref{eq:scalar_equation} and the equations for the mode functions \eqref{eq:mode_function_u}-\eqref{eq:mode_function_v} using the continuum Hamiltonian \eqref{eq:matter_Hamiltonian_continuum}. We note that the mode functions $u_\k(t)$ and $v_{-\k}(t)$ enter the scalar field equation via the definition of the renormalized scalar condensate \eqref{eq:renormalized_scalar_condensate}, which comprises an integral. This integral is evaluated numerically using a rectangle rule. To ensure a fair comparison, we match the IR resolution of the continuum to that of the lattice simulation by using the same momentum spacing $\Delta\k=\frac{\pi}{L}$. On the other hand, we choose a maximum value of the momentum $k_{\rm max}$ such that further increasing it does not modify the evolution of $\phi(t)$ within numerical precision. In particular, it is much larger than the UV cutoff associated to the smallest $a$ studied. This allows to numerically obtain the solution $\phi_{a\to0}(t)$ for $a\to0$, meaning that no lattice artifacts enter the Hamiltonian as we use directly $\hat{H}_{\rm QFT}$ (see Eq.~\eqref{eq:QFT-LFT_connection}) and that the UV cutoff is large enough so convergence has already been reached. Then, we evaluate that solution for the same time points as those in $\{\phi^a_n\}_{n=0}^{N_t}$ and compute the distance using the definition in \eqref{eq:L2_distance_discrete}.

\end{document}